\documentclass[aps,prd,preprint,groupedaddress]{revtex4}
\usepackage{graphicx}

 \newcommand{\nc}{\newcommand}
 \nc{\mb}[1]{\makebox[#1]{}}
 \nc{\CC}{{\scriptscriptstyle CC}}
 \nc{\NC}{{\scriptscriptstyle NC}}
 \nc{\V}{{\rm v}}
 \nc{\W}{{\scriptscriptstyle W}}
 \nc{\X}{{\scriptscriptstyle X}}
 \nc{\Z}{{\scriptscriptstyle Z}}
 \nc{\CS}{{\scriptscriptstyle CS}}
 \nc{\DY}{{\scriptscriptstyle DY}}
 \nc{\PW}{{\scriptscriptstyle PW}}
 \nc{\SB}{{\scriptscriptstyle SB}}
 \nc{\CSV}{{\scriptscriptstyle CSV}}
 \nc{\GLS}{{\scriptscriptstyle GLS}}
 \nc{\CIB}{{\scriptscriptstyle CIB}}
 \nc{\PT}{{\scriptscriptstyle PT}}
 \nc{\IE}{{\it i.e.,\ }}
 \nc{\EG}{{\it e.g.,\ }}
 \nc{\EA}{{\it et al.\ }}
 \nc{\AH}{{\it ad hoc\ }}
 \nc{\CHPT}{{$\chi_{\PT}$\ }}

\nc{\NCA}{{\em Nuovo Cimento}}
\nc{\NIM}{{\em Nucl. Instrum. Methods}}
\nc{\NIMA}{{\em Nucl. Instrum. Methods} A}
\nc{\NPB}{{\em Nucl. Phys.} B}
\nc{\PLB}{{\em Phys. Lett.}  B}
\nc{\PRL}{{\em Phys. Rev. Lett.}}
\nc{\PRD}{{\em Phys. Rev.} D}
\nc{\PRC}{{\em Phys. Rev.} C}
\nc{\ZPC}{{\em Z. Phys.} C}

\nc{\st}{\scriptstyle}
\nc{\sst}{\scriptscriptstyle}
\nc{\mco}{\multicolumn}
\nc{\epp}{\epsilon^{\prime}}
\nc{\vep}{\varepsilon}
\nc{\ra}{\rightarrow}
\nc{\ppg}{\pi^+\pi^-\gamma}
\nc{\nuN}{{\nu N_0}}
\nc{\nubN}{{\overline{\nu} N_0}}
\nc{\snuNC}{{\langle \sigma^{\nuN}_{\NC}\rangle }}
\nc{\snubNC}{{\langle \sigma^{\nubN}_{\NC}\rangle }}
\nc{\snuCC}{{\langle \sigma^{\nuN}_{\CC}\rangle }}
\nc{\snubCC}{{\langle \sigma^{\nubN}_{\CC}\rangle }}
\nc{\Rnu}{{R^{\nu}}}
\nc{\Rnub}{{R^{\overline{\nu}}}}
\nc{\sintW}{{\sin^2 \theta_{\W} }}
\nc{\vp}{{\bf p}}
\nc{\rz}{{1\over \rho_0^2}}
\nc{\ko}{K^0}
\nc{\kb}{\bar{K^0}}
\nc{\al}{\alpha}
\nc{\ab}{\bar{\alpha}}
\nc{\be}{\begin{equation}}
\nc{\ee}{\end{equation}}
\nc{\bea}{\begin{eqnarray}}
\nc{\eea}{\end{eqnarray}}
\begin{document}

\titlepage
%\draft

\title{Charge Symmetry Violating Contributions to Neutrino Reactions} 


\author{J.T.Londergan}
\email{tlonderg@indiana.edu}
\affiliation{Department of Physics and Nuclear
            Theory Center,\\ Indiana University,\\ 
            Bloomington, IN 47405, USA}
%\thanks{}
\author{A.W.Thomas}
\email{athomas@physics.adelaide.edu.au}
\affiliation {Department of Physics and Mathematical Physics,\\ 
                and Special Research Center for the
                Subatomic Structure of Matter,\\ 
                University of Adelaide,
                Adelaide 5005, Australia}
\date{\today}

%\tighten 

\begin{abstract}  
The NuTeV group has measured charged and 
neutral current reactions for neutrinos on iron targets.  
Ratios of these cross sections provide an independent 
measurement of the Weinberg angle. The NuTeV value for $\sintW$ is 
three standard deviations larger than the value measured in other 
electroweak processes. By reviewing theoretical estimates of parton 
charge symmetry violation (CSV), we study CSV contributions to the 
NuTeV measurement. We conclude that 
charge symmetry violating effects should remove roughly 30\% of 
the discrepancy between the NuTeV result 
and other determinations of $\sintW$.   
\end{abstract}

%\narrowtext 
%\twocolumn 

%\keywords{}

\maketitle

\newpage 

%\section{},\subsection{},\subsubsection{}

\section{Introduction\label{Sec:Intro}}

In recent years precise experiments have provided detailed 
information regarding the parton structure of the nucleon.  
One of the more striking developments was the measurement 
of significant differences between up and down antiquark 
distributions in the nucleon sea.  The first clear evidence 
for this was obtained from muon DIS on deuterium by the NMC group 
\cite{NMC,NMCfsv}, that enabled a precise determination of the Gottfried 
Sum Rule \cite{Gottf}.  Later this was more directly confirmed by 
Drell-Yan measurements 
in $pp$ and $pD$ reactions \cite{Na51,E866,E866b} and by 
semi-inclusive electroproduction at HERMES \cite{HERMES}.  
This sea quark flavor asymmetry, which had been 
anticipated \cite{Thomas:1983fh} on the basis
of spontaneously broken chiral symmetry \cite{Thomas:2000ny}, 
has been incorporated 
into the latest phenomenological parton distributions 
\cite{CTEQ5,MRST}.   
  
Another approximate symmetry in parton distributions is charge 
symmetry, which involves rotation of $180^\circ$ about the 
two axis in isospin space, and which corresponds to interchange 
of protons and neutrons (simultaneously interchanging up and down 
quarks).  Our faith in the charge symmetry of parton distributions 
is justified from experience in nuclear physics, where this symmetry 
is respected to a high degree of precision \cite{Miller,Henley}. 
Until recently, the quantitative evidence which could be 
extracted from high energy experiments, although not particularly precise, 
was consistent with charge symmetric parton distributions 
\cite{Lon98}.  As a result, 
all phenomenological analyses of deep inelastic scattering 
data in terms of parton distribution functions assume  
charge symmetry.  Experimental verification of charge symmetry 
was difficult, first because high precision experiments were necessary 
to isolate charge symmetry violating (CSV) effects, second because the 
best tests required 
comparison of electromagnetic and neutrino-induced reactions,  
and third because CSV often mixes with quark flavor asymmetry 
effects.  

Recent precise experiments have now significantly decreased the 
upper limits on parton 
CSV contributions.  The NMC measurements of muon DIS on deuterium 
\cite{NMC} provide values for the charged lepton structure
function $F_2^\mu(x,Q^2)$.  In a similar $Q^2$ regime the 
CCFR Collaboration \cite{CCFR} extracted the structure functions 
$F_2^\nu(x,Q^2)$ from neutrino--induced charge-changing reactions.  
In sec.\ \ref{Sec:Two}, we review the comparison of these structure functions 
and we show that upper limits of a few percent can be placed on 
parton CSV in the region $ 0.01 \le x \le 0.4$. In sec.\ \ref{Sec:Three} 
we review theoretical models for charge symmetry violation 
and we present two models for CSV parton distribution and an 
estimate of CSV parton distributions using phenomenological parton 
distributions.      

Recently, the NuTeV experimental group has measured both neutral 
and charged-current cross sections for neutrinos and antineutrinos 
on iron targets \cite{NuTeV}.  As was originally pointed out 
by Paschos and Wolfenstein \cite{Pas73}, ratios of these cross 
sections on isoscalar targets can provide an independent 
measurement of the Weinberg angle, $\sintW$. The 
value of $\sintW$ extracted by the NuTeV group is 
three standard deviations larger than the measured fit to other 
electroweak processes \cite{EM00}.  This striking result has led some 
people to suggest that they may be seeing evidence 
of physics beyond the Standard Model. 

Because of the importance of this result, it is crucial that 
all the effects contributing to this result be estimated to the best 
of our abilities.  The NuTeV group has recently published 
a paper \cite{NuTeV2} that estimates the contributions to their result 
from three sources: the excess of neutrons in their iron target;  
the possible contribution from strange quarks and the effects 
of parton charge symmetry violation. As we will discuss in the 
following section, the isoscalar correction (excess neutrons in 
iron) makes a rather large, but apparently well-determined, correction to 
the extracted value of the Weinberg angle. At present, even the sign of 
the contribution from strange quarks is uncertain.  In this paper, we 
will show that we can predict the sign of the contribution from parton 
CSV with some confidence.  We will demonstrate that 
different theoretical predictions of the magnitude of the CSV correction 
to the neutrino determination of the Weinberg angle are in 
reasonable agreement with each other.  CSV contributions should decrease 
the discrepancy between this experiment and the value of the 
Weinberg angle extracted from electroweak experiments in 
the vicinity of the $Z$ mass. This will be reviewed in Sec. 
\ref{Sec:Four}.

\section{Extraction of Weinberg Angle from Neutrino Scattering
\label{Sec:Two}}

Paschos and Wolfenstein \cite{Pas73} showed that one could obtain an 
independent measurement of the Weinberg angle, by taking ratios of 
charged-current and neutral-current cross sections for neutrinos 
and antineutrinos on isoscalar targets.  They proposed 
measuring the ratio 
\be
 R^- \equiv { \rz \left( \snuNC - \snubNC \right) \over 
 \snuCC - \snubCC } = {1\over 2} - \sintW
\label{eq:PasW} 
\ee  
In Eq.\ \ref{eq:PasW}, $\snuNC$ is the neutral-current inclusive 
cross section, integrated over $x$ and $y$, for neutrinos on an 
isoscalar target.  The quantity $\rho_0 \equiv M_{\W}/(M_{\Z}\,\cos 
\theta_{\W})$ is one in the Standard Model.  Alternatively, Eq.\ 
\ref{eq:PasW} can be written as 
\bea 
R^- &=& {\left( \Rnu - \Rnub \right) \over 1 - r\, \Rnu} ~,  
 \hspace{0.25truein} r = {\snubCC \over \snuCC} 
 \nonumber \\ \Rnu &=& {\rz \snuNC \over \snuCC}~, \hspace{0.25truein} 
\Rnub = {\rz \snubNC \over \snubCC} . 
\label{eq:PasW2}
\eea
The NuTeV group used the Sign Selected Quadrupole Train beamline at 
FNAL to separate neutrinos and antineutrinos arising from pion 
and kaon decays following the interaction of 800 GeV protons.  The 
resulting interaction events were observed in the NuTeV detector, 
and were required to deposit between $20$ GeV and $180$ GeV in 
the calorimeter.  $CC$ and $NC$ events were distinguished by 
the event length in the counters, as $CC$ events contained a final 
muon that penetrated substantially farther than the hadron shower.  

Rather than working directly with the Paschos-Wolfenstein ratio, the 
NuTeV collaboration measured the individual ratios $\Rnu$ and 
$\Rnub$ defined in Eq.\ \ref{eq:PasW2}, and took the value of $r$ 
from earlier experiments.  From the result, $\Rnu = 0.3916 \pm 
0.0007$ and $\Rnub = 0.4050 \pm 0.0016$, they extracted 
$\sintW = 0.2277 \pm 0.0013 ~(stat) \pm 0.0009 ~(syst)$.  This 
value is three standard deviations above the measured fit to 
other electroweak processes, $\sintW = 0.2227 \pm 0.00037$ \cite{EM00}.  
This result is striking, and if no other effects can account 
for this discrepancy, it may be evidence of physics beyond the 
Standard Model.   

Several approximations have been made in deriving Eq.\ \ref{eq:PasW}.  
It is true only for isoscalar targets,  
includes only the contributions from light quarks, 
and assumes the validity of parton charge symmetry.  The 
NuTeV group has recently investigated how their result changes 
when these assumptions are removed \cite{NuTeV2}. 
The corrections to the Paschos-Wolfenstein ratio of Eq.\ \ref{eq:PasW}
take the form 
\bea 
\Delta R^- &=& {S_{\V} \over U_{\V} + D_{\V}}\left[ 2\Delta_d^2 + 
  3\left(\Delta_u^2 + \Delta_d^2 \right)\epsilon_c \right] \nonumber \\  
 &+& {\left( 3\Delta_u^2 + \Delta_d^2 \right) \over \left(U_{\V} + 
 D_{\V}\right)} \left[ -\delta N \left( U_{\V} - D_{\V} \right) + 
  {1\over 2}\left( \delta U_{\V} - \delta D_{\V} \right) \right] 
  \nonumber \\ Q &\equiv& \int_0^1 x\,q(x)\,dx \nonumber \\ 
 Q_{\V} &\equiv& Q - \overline{Q} 
\label{eq:PWcorr}
\eea
In Eq.\ \ref{eq:PWcorr}, the quantity $Q$ is the total momentum carried 
by a quark of flavor $q$, and the quantity $Q_{\V}$ is the total 
momentum carried by valence quarks of that flavor. $\delta N = (N-Z)/A$ is 
the fractional neutron excess, $\Delta^2_{u,d} = 
(\epsilon^{u,d}_L)^2 - (\epsilon^{u,d}_R)^2$ and $\epsilon_c$ 
is the kinematic suppression factor for massive charm production.  

The isoscalar contribution (the term proportional to $\delta N$ in 
Eq.\ \ref{eq:PWcorr}) is straightforward to calculate. Using the 
values $\delta N = .0567, 3\Delta_u^2 + \Delta_d^2 = .4804$, and 
$(U_{\V} - D_{\V})/(U_{\V} + D_{\V}) \sim 0.46$, gives an 
isoscalar correction to $\sintW$ of about $-0.0125$.  The NuTeV group 
reports an isoscalar correction of $-0.0080$, with a very small error 
\cite{ZelPC}.  This differs from the `naive' correction of  
Eq.\ \ref{eq:PWcorr} because the NuTeV group corrects for factors 
like experimental cuts, experimental backgrounds and 
the enhanced sensitivity of their experiment to neutrino scattering 
at low $x$. All of these factors were used as input in a detailed Monte 
Carlo simulation of their experiment.  However, Kulagin \cite{Kul03} 
has recently claimed that the uncertainties in the isoscalar 
corrections are likely to be considerably larger than estimated by 
the NuTeV group. 

The contribution from strange 
quarks depends on the quantity $S_{\V} \equiv S - \overline{S}$, the 
difference between the momentum carried by strange quarks and strange 
antiquarks.  Even the sign of this quantity is not firmly established.  
Barone \EA \cite{Bar00} analyzed the CDHS neutrino charged-current 
inclusive cross sections and charged lepton structure functions 
\cite{CDHS}, and argued that some improvement is obtained by allowing 
an asymmetric strange sea with $S_{\V} > 0$.  However, the CCFR 
\cite{Yan01} and NuTeV \cite{Gon01} charged-current and dimuon results 
show significant disagreement with the CDHS results at large $x$.  
The NuTeV group finds a best fit to their results with a slightly 
negative value $S_{\V} = -.0027 \pm 0.0013$. 

In the remaining sections, we will review the origin of valence 
quark charge symmetry violation, and we will apply several theoretical 
models to estimate the CSV contribution to the NuTeV value for 
$\sintW$.           
 
\section{Valence Quark Charge Symmetry Violation
\label{Sec:Three}}

Because the Paschos-Wolfenstein relation involves the difference between 
neutrino and antineutrino cross sections, it is dominated by  
contributions from valence parton distributions, 
\be
q_{\V}(x) = q(x) - \overline{q}(x) ~~, 
\label{eq:valq}
\ee
for a particular quark flavor $q$.  
We can gain insight into the origin and magnitude of parton 
charge symmetry violation by using a method for calculating 
twist-two valence parton distributions developed by the Adelaide 
group \cite{Adl90,Lon98}. This method evaluates quark distributions 
through the relation 
\be 
q(x, \mu^2) = M\, \sum_X \, |\langle X | \psi_+(0) | N\rangle |^2 
 \delta( M(1-x) - p_{\X}^+ ) 
\label{eq:qvAdl}
\ee
In Eq.\ \ref{eq:qvAdl}, $\psi_+ = (1+ \alpha_3)\psi/2$ is the 
light front operator that removes a quark or adds an antiquark to 
the nucleon state $|N\rangle$, $\mu^2$ represents the starting 
scale for the quark distribution, $| X\rangle$ are all 
possible final states that can be reached with this operator, 
and $p_{\X}^+$ is the plus component ($p^+ \equiv p_3 + E(p)$) of the 
residual system.  Therefore, $| X\rangle = 2q, 3q + \overline{q}, 4q + 
2\overline{q}, \dots$.  

Thomas and collaborators showed that one could obtain quark 
distributions in semi-quantitative agreement with experiment 
using simple quark models such as the MIT bag \cite{Bag73} in Eq.\ 
\ref{eq:qvAdl}, then evolving the resulting quark distribution from 
the bag scale $\mu^2$ to the final value of $Q^2$. The contribution to 
a quark distribution from intermediate state $X$ produces a peak at a value 
$x_p \sim 1 - M_{\X}/M$, where $M_{\X}$ is the effective mass of 
the state $X$. 
We note that for two-quark intermediate states, $x_p \sim 
1/3$, while $x_p$ is negative for states with four or more quarks. 
Consequently, for $x \ge 0.2$, where valence quarks dominate, 
qualitative estimates can be obtained by including only two-quark 
intermediate states in Eq.\ \ref{eq:qvAdl}. Most importantly, this
observation means that the major
qualitative features of the spin and flavor dependence of the valence parton
distribution functions can be understood simply in terms of the
hyperfine mass splitting between two-quark states with spin zero 
or one \cite{Adl90,Clo88}. A similar analysis can also 
explain the spin and flavor dependence of strange baryon parton 
distributions at large $x$ \cite{Bor00}, and also quark fragmentation 
functions at large energy fraction $z$ \cite{Bor00b}. 

We can use Eq.\ \ref{eq:qvAdl} to estimate charge symmetry 
violating effects, \EG the difference between the up quark valence 
distribution in the proton and the down quark in the neutron, 
$\delta d_{\V}(x) = d_{\V}^p(x) - u_{\V}^n(x)$.  There are four sources of 
CSV contributions: charge symmetry violation in the quark 
wavefunctions; electromagnetic effects that break charge symmetry; 
mass differences of the struck quark; and mass differences in 
the spectator multiquark systems.  Model quark wavefunctions are 
found to be almost invariant under the small mass changes typical 
of CSV.  At sufficiently high energies, electromagnetic effects 
should also be small, and these are neglected.  Consequently, parton 
charge symmetry violation will arise predominantly through mass 
differences $m_d - m_u$ of the struck quark, and from mass differences 
in the spectator multi-quark system.  Both of these contributions 
will result in small shifts in the argument of the energy-conserving 
delta function in Eq.\ \ref{eq:qvAdl}. Since at large $x$ the 
contribution to the parton distribution is dominated by the two-quark 
intermediate state, we can thus obtain quantitative estimates 
of the sign and magnitude of parton CSV from these terms. We stress that
the change in the mass of the spectator pair is
exactly the same mechanism that leads (through a much larger mass difference) 
to an understanding of
the major features of the spin and flavor dependence of  
valence distributions, therefore one has a fair degree of
confidence in this particular term.

Consider the ``minority valence quark'' CSV term, 
\be
 \delta d_{\V}(x) = d_{\V}^p(x) - u_{\V}^n(x) ~~. 
\label{eq:deldv}
\ee
The main contribution to minority valence quark CSV arises from the 
mass of the diquark system, which is $uu$ for the proton and $dd$ for 
the neutron. The two-quark contribution to the valence 
quark distribution will peak at $x_p^p \sim 1- M_{uu}/M$ for 
$d_{\V}^p$ and $x_p^n \sim 1- M_{dd}/M$ for 
$u_{\V}^n$, thus the down quark distribution in the proton will be 
shifted to higher $x$ and the up quark distribution in the neutron 
will be shifted to lower $x$.  This means that, at large $x$, 
$\delta d_{\V}(x)$ will be positive.  

Next, we consider the ``majority valence quark'' CSV term, 
\be
 \delta u_{\V}(x) = u_{\V}^p(x) - d_{\V}^n(x) ~~. 
\label{eq:deluv}
\ee
The mass of the diquark state is $ud$ for both $u_{\V}^p$ and 
$d_{\V}^n$. As a result, the two-quark contribution to the majority 
quark valence distribution will peak at $x_p^p \sim 1- M_{ud}/M^p$ for 
$u_{\V}^p$ and $x_p^n \sim 1- M_{ud}/M^n$ for 
$d_{\V}^n$.  As a result we expect $\delta u_{\V}(x)$ to be negative 
at large $x$.  

An important constraint on CSV parton distributions is that 
they respect the normalization of valence quarks.  The 
integral over $x$ of valence quark distributions must give 
the total number of valence quarks, \IE 
\bea
 \int_0^1 u_{\V}^p(x) \, dx = \int_0^1 d_{\V}^n(x) \, dx &=& 2 
  \nonumber \\ \int_0^1 d_{\V}^p(x) \, dx = \int_0^1 u_{\V}^n(x) \, dx 
  &=& 1 ~~, 
\label{eq:valnum}
\eea
and hence Eq.\ \ref{eq:valnum} requires that 
\be 
\int_0^1 \delta d_{\V}(x) \, dx = \int_0^1 \delta u_{\V}(x) \, dx = 0 . 
\label{eq:delint}
\ee   
It is important that valence quark CSV distributions obey this 
constraint, otherwise one is effectively changing the total number of 
up or down valence quarks in the nucleon.  

\begin{figure}
\vspace{-1.0cm}
\includegraphics[height=10.0cm]{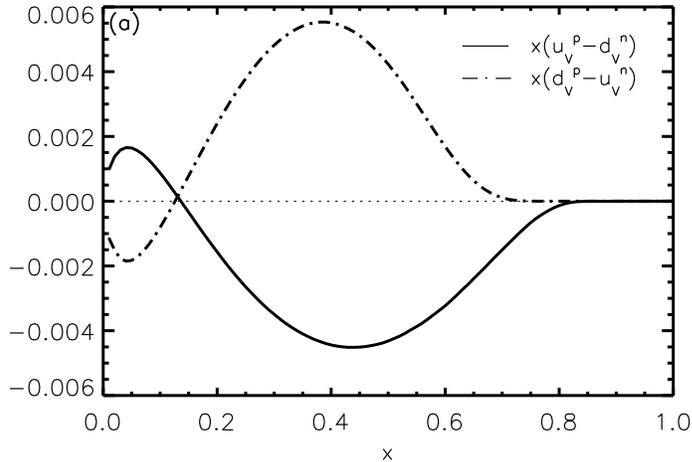}
\vspace{-0.5cm}
\caption{Valence quark CSV contributions, $x\delta q_{\V}(x)$ 
vs.\ $x$. Solid line: $x\delta u_{\V}$; dash-dot line: $x\delta d_{\V}$.  
Calculated using MIT bag model wavefunctions by Rodionov 
\EA, Ref.\ \protect{\cite{Rod94}}, and evolved to $Q^2 = 10$ GeV$^2$.
\label{fig1}}
\end{figure}

Since we can infer the sign and relative magnitude of the CSV 
parton distributions at large $x$, and these distributions must 
integrate to zero, we can obtain at least 
qualitative values for the valence CSV distributions.  $\delta d_{\V}(x)$ 
will be positive at large $x$ and thus negative at small $x$, while  
$\delta u_{\V}(x)$ will be negative at large $x$ and positive at small 
$x$.  In Fig.\ \ref{fig1}, we show $x \delta q_{\V}(x)$ for valence 
up and down distributions as calculated by Rodionov, Thomas and 
Londergan \cite{Rod94}.  These distributions are calculated from 
Eq.\ \ref{eq:qvAdl} using an MIT bag model for the quark 
wavefunctions. These numerical calculations included all four sources of 
CSV noted earlier. The resulting CSV distributions were 
evaluated at 
the bag scale and evolved upwards in $Q^2$.  The distributions 
shown in Fig.\ \ref{fig1} are evolved to $Q^2 = 10$ GeV$^2$. 
The parton distributions exhibit the qualitative effects that 
we inferred; at large $x$, $\delta d_{\V}(x) > 0$ and 
$\delta u_{\V}(x) < 0$.  The CSV distributions must change sign 
at small $x$ to satisfy the requirement of valence quark conservation,
summarized in Eq.\ \ref{eq:delint}.

\begin{figure}
\vspace{-2.0cm}
\includegraphics[height=2.4in]{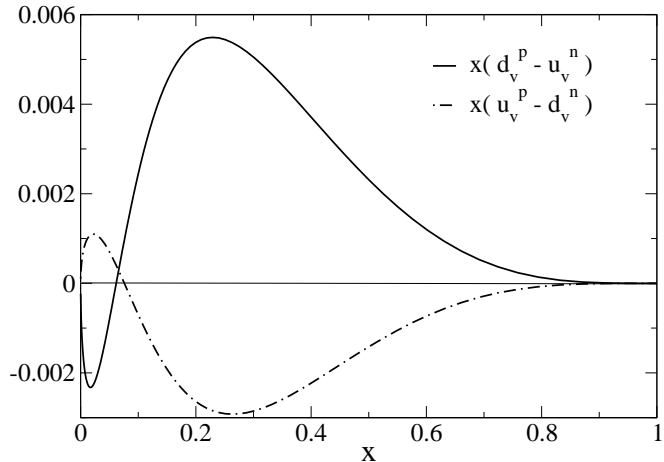}
\vspace{0.3cm}
\caption{Valence quark CSV contributions, $x\delta q_{\V}(x)$ 
vs.\ $x$. Solid line: $x\delta d_{\V}$; dash-dotted line: $x\delta u_{\V}$.  
Calculated from valence quark distributions using the analytic 
approximation of Sather, Ref.\ \protect{\cite{Sat92}}, and the CTEQ4LQ
parton distributions of Ref.\ \protect{\cite{CTEQ3}}, modified 
according to Eq.\ \protect{\ref{eq:CTEQmod}}. These distributions 
were then evolved from the starting scale $Q^2 = 0.49$ GeV$^2$ to 
$Q^2 = 20$ GeV$^2$ appropriate for the NuTeV experiment.
\label{fig3}}
\end{figure}

Charge symmetry violating parton distributions were also calculated 
by Sather \cite{Sat92}. Starting also from Eq.\ \ref{eq:qvAdl}, 
Sather derived the following analytic approximation  
relating CSV distributions to valence parton distributions, 
\bea
 \delta d_{\V}(x) &=& -{\delta M \over M}{d\over dx} \left[ 
 xd_{\V}(x) \right] - {\delta m \over M}{d\over dx} d_{\V}(x)  \nonumber 
 \\ \delta u_{\V}(x) &=& {\delta M \over M}\left(-{d\over dx} \left[ 
 xu_{\V}(x) \right] + {d\over dx} u_{\V}(x) \right) 
\label{eq:Satanl}
\eea
In Eq.\ \ref{eq:Satanl}, $\delta M = M_n - M_p = 1.3$ MeV is the 
neutron-proton mass difference, $\delta m = m_d - m_u$ is the 
down-up quark mass difference, and $M$ is the average nucleon mass. 
To calculate the CSV parton distributions, Sather used parton 
distributions obtained from the MIT bag model in Eq.\ \ref{eq:Satanl}.  
Qualitatively, Sather's CSV parton distributions are  
quite similar to those of Rodionov \EA \cite{Rod94}.  However, 
$\delta u_{\V}(x)$ is smaller than the corresponding quantity for 
Rodionov \EA, and Sather's CSV distributions peak at somewhat smaller 
$x$. Sather's CSV distributions 
are qualitatively similar to those shown in Fig.\ \ref{fig3}.      

Both of these charge symmetry violating distributions were 
constructed using bag model wavefunctions.  Such wavefunctions can 
give qualitative agreement with phenomenological parton distributions, 
but these model-generated parton distributions systematically underpredict
the results of phenomenological distributions at large $x$.  Here, we 
present a 
method for generating valence quark CSV parton distributions from 
phenomenological distributions.  We will use the approximate formula 
derived by Sather \cite{Sat92}, Eq.\ \ref{eq:Satanl}, that relates CSV 
parton distributions 
to the derivatives of valence quark parton distributions. 

Although Sather obtained his CSV distributions by using parton 
distributions from the MIT bag model in Eq.\ \ref{eq:Satanl}, we 
will insert phenomenological parton distributions into this equation.  
There is a problem with this approach.  As we pointed 
out in Eq.\ \ref{eq:delint}, conservation of valence quark probability 
requires that the integral over all $x$ of $\delta q_{\V}$ in 
Eq.\ \ref{eq:Satanl} be zero.  Since the terms in this equation are 
just derivatives of parton distributions, integration over $x$ simply 
involves evaluating the parton distributions at zero and one.  However, 
phenomenological parton distributions go like $x^{-1/2}$ in the 
limit $x \ra 0$, hence when phenomenological distributions 
are used the integral of the CSV distributions 
in Eq.\ \ref{eq:Satanl} will not be zero, in fact the integrals will blow 
up.  

The problem originates because Eq.\ \ref{eq:Satanl} is a reasonable 
approximation for the parton distribution that arises when a nucleon 
consisting of three valence quarks splits into a quark and 
diquark.  This gives the dominant contribution to the parton distribution at 
large $x$.  However, at small $x$ the valence distribution is 
dominated by higher mass Fock states that include many 
quark-antiquark pairs.  For states involving such large excitations, 
the effects of quark and nucleon mass 
differences should be negligible.  However, Eq.\ \ref{eq:Satanl} 
incorrectly predicts very large contributions at small $x$.  
Consequently, we need to suppress the large CSV effects produced 
by Eq.\ \ref{eq:Satanl} at small $x$.  We will deal with this 
problem in a simple, and completely phenomenological, way.  
Parton distributions from the CTEQ collaboration \cite{CTEQ3,CTEQ4} have 
been parameterized using the form 
\be
 q_{\V}(x) = N \left[ x^{\alpha}+ Cx^{\gamma} \right](1-x)^{\beta}  
\label{eq:CTEQqv}
\ee
The small-$x$ behavior is governed by the parameter $\alpha \sim 
-0.5$.  We want a modified 
parton distribution that will vanish at small $x$,  
leaving the large-$x$ behavior relatively unchanged.  We thus replace  
the CTEQ parton distributions $q_{\V}(x)$ in Eq.\ \ref{eq:Satanl} with 
$\widetilde{q}_{\V}(x)$, defined by    
\be
 \widetilde{q}_{\V}(x) = N \left[ x^{\alpha}+ Cx^{\gamma} \right]
\left[ (1-x)^{\beta}- (1-x)^{\beta+12}\right]   
\label{eq:CTEQmod}
\ee
By inspection, $\overline{q}_{\V}(x)$ vanishes at $x=0$, and for 
large  $x$ there is a very small difference between the modified 
parton distribution 
of Eq.\ \ref{eq:CTEQmod} and the phenomenological parton 
distribution. 

We calculated valence quark CSV distributions using Eq.\ 
\ref{eq:Satanl} with the CTEQ4LQ phenomenological parton 
distributions \cite{CTEQ4}, modified using Eq.\ \ref{eq:CTEQmod}.  
The coefficients for the CTEQ4LQ parton distributions, appropriate for 
$Q^2 = 0.49$ GeV$^2$, are given in Table \ref{Tab1}.   
      
\begin{table}%[H] add [H] placement to break table across pages
\caption{\label{Tab1}Parameters for valence quark distribution from CTEQ4LQ 
parton distribution, Ref.\ \protect\cite{CTEQ4}.}
\begin{ruledtabular}
\begin{tabular}{lccccc}
% Lines of table here ending with \\
 & N & $\alpha$ & $\beta$ & C & $\gamma$ \\ \hline
$u_{\V}$ & 1.315 & -0.427 & 3.281 & 10.614 & 0.607 \\
$d_{\V}$ & 0.852 & -0.427 & 4.060 & 4.852 & 0.266 \\
\end{tabular}
\end{ruledtabular}
\end{table}

Sather's analytic expression, Eq.\ \ref{eq:Satanl}, is appropriate 
for a nucleon at a low starting scale, of order $Q^2 \sim 0.25 - 
0.5$ GeV$^2$.  The CTEQ4LQ parton distribution \cite{CTEQ4} was 
introduced specifically for starting scale $Q^2 = 0.49$ GeV$^2$.  
The resulting CSV distributions were evolved to the higher value 
$Q^2 = 20$ GeV$^2$ 
appropriate for the NuTeV experiment, using the QCD evolution 
program of Miyama and Kumano \cite{Miy96}.  The evolved CSV parton 
distributions are shown in Fig.\ \ref{fig3}.  The solid curve 
shows $x\delta d_{\V}$ and the dash-dotted curve gives $x\delta u_{\V}$.   

\section{Charge Symmetry Violating Contributions to Neutrino Reactions
\label{Sec:Four}}

Using the valence quark CSV distributions reviewed in Sec.\ 
\ref{Sec:Three}, we can estimate the CSV contribution to the extracted 
value of $\sintW$.  From Eq.\ \ref{eq:PWcorr}, the CSV corrections to the 
Paschos-Wolfenstein ratio are of the form 
\bea
\Delta R^-_{\CSV} &=& \left( 3\Delta_u^2 + \Delta_d^2 \right) 
  {\delta U_{\V} - \delta D_{\V} \over 2(U_{\V} + D_{\V})} 
  \equiv \Delta U_{\V} + \Delta D_{\V}  \nonumber \\ 
  Q_{\V} &=& \int_0^1 \,x\, q_{\V}(x) \,dx \nonumber \\ 
  \delta D_{\V} &=& \int_0^1 \,x\, \left[ d_{\V}^p(x)- u_{\V}^n(x)
   \right]\,dx  \nonumber \\ 
  \delta U_{\V} &=& \int_0^1 \,x\, \left[ u_{\V}^p(x)- d_{\V}^n(x)
   \right]\,dx  
\label{eq:PWCSV}
\eea           

In Table \ref{Tab2} we show the contributions of the different 
theoretical CSV estimates to the Paschos-Wolfenstein ratio.  We have 
broken down the individual contributions so that $\Delta U_{\V}$ and 
$\Delta D_{\V}$ are the total CSV effects arising from $\delta U_{\V}$ 
and $\delta D_{\V}$, respectively. From Fig.\ 
\ref{fig1}-\ref{fig3}, in all cases both $\delta U_{\V}$ and $\delta D_{\V}$ 
make a negative contribution.  
Therefore the net CSV contribution will be 
negative.  This will decrease the discrepancy between the value of 
$\sintW$ extracted in the NuTeV experiment, and the best value 
obtained from high-energy electroweak interactions. 

\begin{table}%[H] add [H] placement to break table across pages
\caption{\label{Tab2}CSV corrections to determination of $\sintW$ 
in neutrino scattering. $PW$ is contribution to the 
Paschos-Wolfenstein ratio, $Nu$ is weighted by the NuTeV functional.  
$\Delta U$ ($\Delta D$) is total contribution from 
$\delta u_{\V}$ ($\delta d_{\V}$), and $Tot$ 
is total CSV correction.}
\begin{ruledtabular}
\begin{tabular}{lcccccc}
% Lines of table here ending with \\
 & $\Delta U_{\PW}$ & $\Delta D_{\PW}$ & $Tot_{\PW}$ & $\Delta U_{Nu}$ & 
 $\Delta D_{Nu}$ & $Tot_{Nu}$ \\ \hline
Rodionov & -.0010 & -.0011 & -.0020 & -.00065 & -.00081 & -.0015 \\
Sather & -.00078 & -.0013 & -.0021 & -.00060 & -.0011 & -.0017 \\
analytic & -.00075 & -.0013 & -.0021 & -.0005 & -.0009 & -.0014 \\
\end{tabular}
\end{ruledtabular}
\end{table}

Table \ref{Tab2} shows that CSV contributions to the 
Paschos-Wolfenstein ratio from three theoretical models 
give almost identical results,  
$-0.0020$ or $-.0021$.  Since the value of $\sintW$ from the 
NuTeV experiment, without CSV corrections, was $0.005$ larger than 
the best value from electromagnetic interactions,  
our calculated CSV effect would reduce the discrepancy between 
the neutrino and electromagnetic measurements of $\sintW$ by 
$40$\%.  

However, as pointed out by the NuTeV group \cite{NuTeV2}, it is 
not appropriate to use Eq.\ \ref{eq:PWCSV} to determine the 
CSV effects on the value of $\sintW$, as the NuTeV extraction 
of this quantity does not rely on the Paschos-Wolfenstein ratio, 
but instead uses the absolute ratios $\Rnu$ and $\Rnub$ defined 
in Eq.\ \ref{eq:PasW2}, and compares the results with a full Monte 
Carlo simulation of the experimental processes.  The NuTeV group 
produced functionals that give the sensitivity of their observables 
to various effects.  These are summarized in a single integral, 
\be
  \Delta {\cal E} = \int_0^1 \, F\left[ {\cal E}, \delta; x\right] 
  \, \delta(x) \, dx
\label{eq:Func}
\ee
Eq.\ \ref{eq:Func} gives the change in the extracted quantity 
${\cal E}$ resulting from the symmetry violating quantity $\delta(x)$.  
The functionals appropriate for the observable $\sintW$ and 
the parton CSV distributions, were provided in Ref.\ \cite{NuTeV2}.  

We have taken the parton CSV distributions and folded them 
with the NuTeV functionals.    
The net CSV correction to the value of $\sintW$ is 
listed as $Tot_{Nu}$ in Table \ref{Tab2}.  
The CSV contributions are still negative, \IE 
they decrease the discrepancy between the neutrino and 
electromagnetic measurements of the Weinberg angle.  However, 
the CSV change in $\sintW$ is slightly smaller than 
estimated in Eq.\ \ref{eq:PWCSV}.  This is because the NuTeV 
experiment is somewhat more sensitive to small-$x$ physics than 
to effects at larger $x$.  The CSV parton distributions, weighted 
by $x$, change sign at small $x$ and reach a maximum 
at larger $x$.  Thus the functionals, which emphasize 
the CSV distributions at small $x$ where they are small and change 
sign, and de-emphasize them at larger $x$, 
give a smaller CSV effect than predicted by the Paschos-Wolfenstein 
ratio.  Nevertheless, the CSV contributions 
to the NuTeV result, weighted with their functionals, range from 
$-0.0014$ to $-.0017$, and thus reduce the discrepancy between 
the neutrino and electromagnetic measurements of $\sintW$ by 
roughly $30$\%. 

The NuTeV group estimated the CSV correction \cite{NuTeV2} and obtained 
a much smaller value than ours.  However, in order to obtain CSV 
parton distributions, they took the ratio 
$\delta q_{\V}(x)/q_{\V}(x)$ from Rodionov \EA \cite{Rod94}, and 
multiplied this ratio by parton distributions determined from 
neutrino scattering.  Since the ratio was determined using parton 
distributions from a quark model, and the parton distributions were 
obtained from quite a different source, we are not convinced of the 
accuracy of the resulting CSV distributions.  In particular, 
CSV distributions constructed in this way will not satisfy any 
relation such as Eq.\ \ref{eq:Satanl}, nor will they satisfy 
valence quark conservation of Eq.\ \ref{eq:delint}.        
    
\section{Conclusions\label{Sec:Concl}} 

We have reviewed the corrections that parton charge symmetry violation 
should make in determining $\sintW$ in neutrino 
scattering.  Although parton CSV effects are sufficiently small that 
neither their sign nor magnitude has been measured to date, we argued 
that at large $x$ the CSV distribution 
$\delta d_{\V}(x)$ should be positive and $\delta u_{\V}(x)$ should be 
negative.  The CSV distributions must preserve the overall 
number of valence quarks in the neutron, hence from Eq.\ \ref{eq:delint} 
both CSV distributions must change sign at small $x$. 

We investigated two theoretical models of parton CSV.  In both cases, 
the parton distributions are calculated from a model of QCD, such as 
the MIT bag, and the models determined how the CSV distributions 
could be related to the parton distributions and their derivatives.  
In a third case, we inserted phenomenological parton distributions
from the CTEQ group into analytic expressions relating the CSV 
distributions to parton distributions.  We removed the tendency 
of these analytic expressions to (incorrectly) 
give too much weight to the small-$x$ region, by damping the 
phenomenological distributions at very small $x$. In all cases, the 
CSV distribution was obtained at a low starting scale, and then 
evolved to higher $Q^2$, more appropriate for the $Q^2$ value of 
the neutrino experiments. 

All of these theoretical CSV distributions serve to decrease the 
value of $\sintW$ extracted from the NuTeV experiment; the 
size of these corrections was remarkably similar, ranging from 
$-.0014$ to $-.0017$. Since the 
value of $\sintW$ extracted from neutrino scattering was greater than 
that obtained from electromagnetic interactions by $+.005$, inclusion 
of a charge symmetry violating contribution of this magnitude would reduce 
this discrepancy by about 30\%.  We emphasize that CSV effects 
have yet to be confirmed by direct experiment, so our calculations 
only give the best theoretical estimate of these contributions.  
The predicted CSV effects arise from the same mechanism that correctly 
predicts the spin and flavor dependence of valence quark distributions, 
so we would be surprised if our 
CSV effects did not have the correct sign and roughly the right 
magnitude.  It is rather remarkable that although 
the predicted corrections from parton charge symmetry violation are 
quite small, high energy 
experiments have now reached a precision where even small 
absolute effects have a significant impact on our ability to 
extract fundamental quantities.

\vspace*{0.3cm}
\begin{acknowledgments}
This work was supported in part by the Australian Research Council.  
One of the authors [JTL] was supported in part by the National
Science Foundation research contract PHY--0070368.  The authors wish 
to thank G.P. Zeller and K. McFarland of the NuTeV collaboration 
for very useful comments regarding the NuTeV measurements and 
calculations, and about CSV contributions to extraction of the 
Weinberg angle.  JTL wishes to acknowledge discussions with 
W. Melnitchouk and C. Benesh.  
\end{acknowledgments}

\end{document}